\documentclass[
aps,prd,
12pt,%10pt
nopreprintnumbers,
showpacs,
eqsecnum,
nofootinbib
]{revtex4-1}

\usepackage{graphicx}
\usepackage{amssymb}

\begin{document}

\title{Calculations in induced gravity from higher-derivative f{}ield theories}
\author{Nahomi Kan}\email[]{kan@gifu-nct.ac.jp}
\affiliation{National Institute of Technology, Gifu College,
Motosu-shi, Gifu 501-0495, Japan
}
%\author{Masashi Kuniyasu}\email[]{mkuni13@yamaguchi-u.ac.jp}
\author{Kiyoshi Shiraishi}\email[]{shiraish@yamaguchi-u.ac.jp}
%\author{Kohjiroh Takimoto}\email[]{i016vb@yamaguchi-u.ac.jp}
%\author{Zhenyuan Wu}\email[]{b501wb@yamaguchi-u.ac.jp}
\affiliation{
Graduate School of Sciences and Technology for Innovation, Yamaguchi
University, Yamaguchi-shi, Yamaguchi 753--8512, Japan}
\date{\today}
%\date{}

\begin{abstract}
In this paper, we investigate Einstein's gravity induced from 
higher-derivative scalar field theories. We develop an approach utilizing an
effective theory of multiple fields for the higher-derivative theory. The
expressions for induced cosmological constant and the induced gravitational
constant are obtained in the present scenario of induced gravity in $D$
dimensions. We also show that finite values for the induced constants can be
extracted in certain infinite-derivative theories.
\end{abstract}

%\preprint{}

\pacs{%
%02.10.Ox, %%%Combinatorics; graph theory
%02.20.Sv, %Lie algebra of Lie groups
%02.30.Hq, %Ordinary differential equations
%02.30.Ik, %Integrable systems
%02.30.Jr, %Partial differential equations
%02.40.Gh, %Noncommutative geometry
%03.65.-w, %Quantum mechanics
%03.65.Db, %Functional analytical methods
%03.65.Sq, %Semiclassical theories in quantum mechanics
%03.70.+k, %Theory of quantized fields
%04.20.-q, %%%Classical general relativity
%04.20.Fy, %%Canonical formalism, Lagrangians, and variational principles
%04.20.Jb, %%Exact solutions
%04.25.-g, %Approximation
%04.25.Nx, %%%Post-Newtonian approximation; perturbation theory; related
%approximations
%04.40.-b, %Self-Gravitating systems
%04.40.Nr, %%Einstein-Maxwell spacetime
%04.50.-h, %%%%%Higher-dimensional gravity and other theories of gravity 
%04.50.Cd, %Kaluza-Klein theories 
%04.50.Gh, %Higher-dimensional black holes, black strings, 
%and related objects 
04.50.Kd, %%%Modified theories of gravity 
%04.60.-m, %%Quantum gravity
%04.60.Kz, %%Lower dimensional models; minisuperspace models
%04.60.Rt, %Topologically massive gravity
04.62.+v, %Quantum fields in curved spacetime
%04.65.+e, %Supergravity
%04.70.Bw, %%%Classical black holes
%05.30.Jp, %Boson systems
%11.10.-z, %%%Field theory
%11.10.Lm, %%%Nonlinear or nonlocal theories and models 
%11.10.Nx, %%%Noncommutative field theory 
%11.10.Kk %%%Field theories in dimensions other than four
%11.25.-w, %Strings and branes
%11.25.Mj, %%Compactification and four-dimensional models
%11.27.+d% %%Extended classical solutions; cosmic strings, 
%domain walls, texture 
%11.30.-j, %Symmetry and conservation laws
%11.30.Pb, %Supersymmetry
11.90.+t   %Other topics in general theory of fields and particles
%12.60.-i, %Models beyond the standard model
12.90.+b   %Miscellaneous theoretical ideas and models
%45.20.Jj, %Lagrangian and Hamiltonian mechanics
%95.35.+d, %Dark matter
%95.36.+x, %Dark energy
%98.80.-k, %%%Cosmology 
%98.80.Cq, %%%%%Particle-theory and field-theory models of the early
%Universe
%98.80.Dr, %Relativistic cosmology 
%98.80.Qc, %Quantum cosmology
%98.80.Jk% %%Mathematical and relativistic aspects of cosmology
}

\maketitle

%%%%%%%%%%%%%%%%%%%%%%%%%%%%%%%%%%%%%%%%%%%%%%%%%%%%%%%%%%%%%%%%%%%%%%%%%%%
%%%%%%%%%%%%%%%%%%%%%%%%%%%%%%%%%%%%%%%%%%%%%%%%%%%%%%%%%%%%%%%%%%%%%%%%%%%
%%%%%%%%%%%%%%%%%%%%%%%%%%%%%%%%%%%%%%%%%%%%%%%%%%%%%%%%%%%%%%%%%%%%%%%%%%%
\section{Introduction}
\label{introduction}
%%%%%%%%%%%%%%%%%%%%%%%%%%%%%%%%%%%%%%%%%%%%%%%%%%%%%%%%%%%%%%%%%%%%%%%%%%%
%%%%%%%%%%%%%%%%%%%%%%%%%%%%%%%%%%%%%%%%%%%%%%%%%%%%%%%%%%%%%%%%%%%%%%%%%%%
%%%%%%%%%%%%%%%%%%%%%%%%%%%%%%%%%%%%%%%%%%%%%%%%%%%%%%%%%%%%%%%%%%%%%%%%%%%

In quantum field theory, the issue of ultraviolet (UV) divergence has been
repeatedly discussed by many authors. In particular, various models in quantum
theory including the gravitational field have been proposed as approaches to
avoid the UV divergence, including attempts beyond the field theory. 

The simplest idea is seen in a
scenario where the theory with higher derivative of fields manages to improve 
the power counting for the UV
divergences \cite{PU,LW1,LW2,LW3,CLP,GOW}. This is owing to the milder behavior of
the Green's function%
\footnote{We use the term ``Green's function'' instead of ``propagator'', because
we use the Euclidean metric in this paper.} at a short distance in the
higher-derivative theory than that in the canonical field theory. It is recently
found that the vacuum expectation value of the scalar field squared in a certain
higher-derivative theory can be reduced to be finite and free from divergence at a
short distance from a conical spatial defect
\cite{KKSW1}, whereas contrary to naive expectations, the quantum fluctuation of
the stress tensor in the higher-derivative theory still suffers from UV divergences
\cite{KKS2020}. Although higher-derivative theories also have a difficult problem
of causality
\cite{Nakanishi,BG,AP,Anselmi1,Anselmi2} in addition, it is worth studying for 
effectiveness of higher derivatives in resolving general UV behavior of quantum
field theory of higher-order gravity \cite{Stelle}%
\footnote{The authors of Refs.~\cite{OS1,OS2,SC,EOS}
considered induced gravity from the phase transition in the theory
with higher derivatives and non-minimally coupled scalar fields. The author of
Ref.~\cite{Narain} considered induced gravity from higher-order gravity with
matter fields. Note that the Adler--Zee formulas (\ref{AZ1}) and (\ref{AZ2}) cannot
be applied to these approaches in their original form.} and that with infinite
derivatives \cite{Modesto1,BT} and arbitrarily high
derivatives \cite{ALS,MS,Modesto2}.

There is a concept of induced
gravity
\cite{Sakharov,Adler1,Adler2,Adler3,Zee1,Zee2,MW,DS,FF,%
Visser,BS,KS1,KS2,KKS,DM,Azri}
as a completely different prospective description of the theory of gravity. It is
based on the idea that the effective theory of the gravitational field is derived
from the quantum effect of matter fields. However, because the gravitational
constant has a dimension of the inverse square of mass in four dimensions, it is
affected by the quadratic divergence in quantum field theory, so many models
cannot predict even a fixed amount of the induced gravitational constant
\cite{Adler2,Visser}. The estimation of the induced cosmological constant resulting
from quantum effects, which is order of the quartic divergence, is more difficult
to compare with the astrophysical knowledge \cite{Padmanabhan}. 

In this paper, we consider the gravity induced from the
quantum theory with higher derivatives of a field. As already mentioned above,
higher derivatives do not necessarily suppress divergences which appear in
stress tensors. Our interest in higher-derivative theory lies in the other two
features of the theory.
First, any higher-derivative theory has a fundamental length scale, as seen from
the dimensional counting. Second, higher-derivative theories are expressed by
multiple fields. In higher-derivative scalar field theory, it is
already known that the degree of freedom of a field increases as the number of
derivatives increases \cite{GOW,CL,Carone,CK,KLS,GPS}.

Recently, Kehagias \textit{et al.}\cite{KPV} investigated induced gravity
in higher dimensional theories not only by the heat kernel method
\cite{DS,FF,Visser,BS,KS1,KS2,KKS,Vassilevich} but
also by the original methods \cite{Adler1,Adler2,Adler3,Zee1,Zee2,MW,DM}.
In their model, infinitely many excited states
\textit{\`a la} Kaluza--Klein theory play an important role in calculability of the
cosmological constant and the gravitational constant without ambiguities.%
\footnote{A similar scenario has been demonstrated earlier by using
dimensional deconstruction \cite{KS1,KS2,KKS} with the method of heat kernel. The
condition for cancellation in UV divergences is identical to what was stated by
Frolov and Fursaev \cite{FF}.} In their model, the dimensionful constants
are proportional to the appropriate powers of the compactification scale.
Therefore, we come to the idea that the induced gravity from higher-derivative
field theories also may possess predictability in gravitational physics at least in
a lowest order of perturbative quantum effects.

In this paper, we present a new model of 
induced gravity, which arises from the quantum effects in
higher-derivative scalar field theory. 
We emphasize that calculable examples of induced gravity can be provided with
certain higher-derivative theories. 
%R
Interestingly, the calculations in those examples result in similar calculations
to those in the model of Ref.~\cite{KPV} based on the Kaluza--Klein theory.
%R
A discussion of unitarity and causality falls
outside the scope of this paper. 

The main results we find are obtained by early
standard methods of Adler--Zee formula \cite{Adler1,Adler2,Adler3,Zee1,Zee2,MW,DM}.
We denote the induced gravitational action as
\begin{equation}
S=\int d^Dx \sqrt{-g}
\frac{1}{16\pi G_{ind}}\left(R-2\Lambda_{ind}\right)\,,
\end{equation}
where $R$ is the scalar curvature. Then, the induced cosmological constant
$\Lambda_{ind}$ is given by
\begin{equation}
\frac{\Lambda_{ind}}{8\pi G_{ind}}=-\frac{1}{D}\langle T(0)\rangle\,,
\label{AZ1}
\end{equation}
and the formula yields the induced gravitational constant $G_{ind}$ as
\begin{equation}
\frac{1}{16\pi G_{ind}}=-\frac{1}{4D(D-1)(D-2)}
\int d^Dw\, |w|^2 \langle \bar{T}(x) \bar{T}(y)\rangle\,,
\label{AZ2}
\end{equation}
where $w\equiv x-y$, the brackets $\langle~\rangle$ indicate a vacuum expectation
value, and
$T\equiv T_{\mu}^{\mu}$ is the trace of the stress tensor of matter fields and
$\bar{T}\equiv T-\langle T\rangle$. It should be noted that we are working in
$D$-dimensional Euclidean space
\cite{DM}. In the following sections, we will apply the formulas above
to the higher-derivative scalar field theories.
%R
A brief review of the derivation of (\ref{AZ1}) and (\ref{AZ2}) can be found in
Appendix \ref{AZR}.

This paper is organized as follows.
The Section \ref{sec2} gives a brief overview of higher-derivative theory and its
effective action with multiple fields. The stress tensor of the theory is obtained
here and the induced cosmological and gravitational constants are
formulated by using the expression of the stress tensor. Section \ref{sec3} deals
with a case study of a certain higher-derivative theory with an infinite number of
derivatives. Some conclusions are drawn in the final section.

%%%%%%%%%%%%%%%%%%%%%%%%%%%%%%%%%%%%%%%%%%%%%%%%%%%%%%%%%%%%%%%%%%%%%%%%%%%
%%%%%%%%%%%%%%%%%%%%%%%%%%%%%%%%%%%%%%%%%%%%%%%%%%%%%%%%%%%%%%%%%%%%%%%%%%%
%%%%%%%%%%%%%%%%%%%%%%%%%%%%%%%%%%%%%%%%%%%%%%%%%%%%%%%%%%%%%%%%%%%%%%%%%%%
\section{induced gravity from the effective Lagrangian and the stress tensor}
\label{sec2}
%%%%%%%%%%%%%%%%%%%%%%%%%%%%%%%%%%%%%%%%%%%%%%%%%%%%%%%%%%%%%%%%%%%%%%%%%%%
%%%%%%%%%%%%%%%%%%%%%%%%%%%%%%%%%%%%%%%%%%%%%%%%%%%%%%%%%%%%%%%%%%%%%%%%%%%
%%%%%%%%%%%%%%%%%%%%%%%%%%%%%%%%%%%%%%%%%%%%%%%%%%%%%%%%%%%%%%%%%%%%%%%%%%%

Recently, Gibbons \textit{et al.}\cite{GPS} proposed an effective Lagrangian of
multiple fields for a general higher-derivative scalar field theory.
They considered the Lagrangian with an arbitrary number of d'Alembert operators
$\Box$ on a real scalar field $\phi$:
\begin{equation}
\mathcal{L}=-\frac{1}{2C}\phi(x)\prod_{i=1}^nA_{i}\phi(x)\,,
\label{ol}
\end{equation}
where $A_i\equiv-\Box+m_i^2$ and $C$ is a constant.
The effective Lagrangian corresponding this Lagrangian reads \cite{GPS}
\begin{equation}
\mathcal{L}=-\frac{1}{2C}\sum_{k=0}^n\eta_k(x)A_{k+1}\chi_{k+1}(x)
+\frac{1}{2C}\sum_{k=1}^n\eta_k(x)\chi_k(x)
\,,
\label{el}
\end{equation}
where $\eta_0\equiv\phi$ and $\chi_n\equiv\phi$. We obtain the
relations, $\eta_k=\left[\prod_{i=1}^kA_i\right]\phi~(1\le k\le n)$, and $\chi_k=
\left[\prod_{i=k+1}^n A_i\right]\phi~(1\le k\le n-1)$ from the iterative use of
the equations of motion. If the expression for $\eta_i$ and $\chi_i$ constructed
from $\phi$ is substituted into (\ref{el}), the original Lagrangian (\ref{ol}) is
recovered.

Then, the stress tensor of the theory is given by \cite{GPS}
\begin{equation}
T_{\mu\nu}=\frac{1}{C}\sum_{k=0}^n\left[
\partial_{(\mu}\eta_k\partial_{\nu)}\chi_{k+1}-\frac{1}{2}g_{\mu\nu}
(\partial_{\rho}\eta_k\partial^{\rho}\chi_{k+1}+m_{k+1}^2\eta_k\chi_{k+1})\right]
+\frac{1}{2C}g_{\mu\nu}\sum_{k=1}^n\eta_k\chi_k\,,
\label{Tmn}
\end{equation}
where $g_{\mu\nu}$ is the metric tensor of the background (flat) $D$
dimensional spacetime. At a glance, it seems inconsistent with the effective
description of the Lee--Wick scalar field theory \cite{GOW} with apparently
indefinite signs of kinetic terms. This is consistent, however, as illustrated in
Appendix \ref{LeeW} for $n=2$.

Now, returning to (\ref{Tmn}), the trace of the stress tensor $T\equiv T_\mu^\mu$
can be written by
\begin{equation}
T=\frac{1}{C}\sum_{k=0}^n\left[-\frac{D-2}{2}
\partial_{\rho}\eta_k\partial^{\rho}\chi_{k+1}-\frac{D}{2}m_{k+1}^2\eta_k
\chi_{k+1}\right]
+\frac{D}{2C}\sum_{k=1}^n\eta_k\chi_k\,.
\label{TT}
\end{equation}
We use this form of the trace of the stress tensor in the formulas
(\ref{AZ1}) and (\ref{AZ2}). Further, we only use the two-point functions such as
$\langle\eta_i(x)\chi_j(y)\rangle$, which is evaluated from the Green's function
$\langle\phi(x)\phi(y)\rangle$. Then, we regard all fields not
independent of $\phi$. We can simplify (\ref{TT}) as%
\footnote{Here, the d'Alembertian should be considered to act on both $\eta_k$ and
$\chi_{k+1}$ symmetrically.}
\begin{equation}
T=\frac{1}{C}\sum_{k=0}^n\left[-\frac{D-2}{2}
\partial_{\rho}\eta_k\partial^{\rho}\chi_{k+1}-\frac{D}{2}\eta_k\Box
\chi_{k+1}\right]\,,
\label{TT2}
\end{equation}
where we used
$\chi_k=\left[\prod_{i=k+1}^n A_i\right]\phi=A_{k+1}\chi_{k+1}$.
The essential advantages of this form (\ref{TT2}) is that we have only to
concentrate ourselves on evaluations of vacuum expectation values of fields since
the contributions of masses are implicit in the expression.

To obtain the quantum quantities, the fundamental basis we use is the Green's
function
\begin{equation}
\langle\phi(x)\phi(x')\rangle=G(x,x')=\frac{C}{\prod_{i=1}^n A_i}\mathbf{1}_{xx'}
\,,
\end{equation}
where $\mathbf{1}_{xx'}$ denotes a covariant delta function
$\frac{1}{\sqrt{|g|}}\delta^D(x,x')$ in this symbolic expression.
From the Green's function and the relations to the original scalar field $\phi$, it
turns out to be
\begin{equation}
\langle\eta_k(x)\chi_{k+1}(x')\rangle=
\left[\prod_{i=1}^kA_i\right]_x\left[\prod_{i=k+2}^n
A_i\right]_{x'}\langle\phi(x)\phi(x')\rangle=\frac{C}{
A_{k+1}}\mathbf{1}_{xx'}=C\Delta_{k+1}(x,x')
\,,
\end{equation}
where $\Delta_{k+1}(x,x')$ is the Green's function of a canonical scalar field
with mass $m_{k+1}$. 
This result is consistent if we consider $\eta$'s and $\chi$'s form a set of free
fields governed by the Lagrangian (\ref{el}). Note that the induced cosmological
and gravitational constants are evaluated by the Adler--Zee formulas with these
Green's functions.

Consequently, we find a fairly simple expression for the induced cosmological
constant
\begin{equation}
\frac{\Lambda_{ind}}{8\pi G_{ind}}=-\frac{1}{D}\langle T\rangle=\lim_{x'\rightarrow
x}\sum_{k=1}^n\left[\frac{D-2}{2D}
\partial_{\lambda}\partial^{\lambda'}+\frac{1}{2}\Box_x\right]\Delta_k(x,x')
\,.\label{cosc}
\end{equation}

On the other hand, the application of the Adler--Zee formula for the induced
gravitational constant to the present theory yields (for a proof, see Appendix
\ref{prf})
\begin{eqnarray}
\frac{1}{16\pi G_{ind}}&=&-\frac{1}{4D(D-1)(D-2)}\int d^Dw\, |w|^2\nonumber \\
& &\times 2\sum_{k=1}^n\left[
\frac{(D-2)^2}{4}\partial_\rho\partial_{\sigma'}\Delta_k(x,x')
\partial^\rho\partial^{\sigma'}\Delta_k(x,x')+\frac{D(D-2)}{2}
\partial_{\rho}\Delta(x,x')\partial^{\rho}\Box\Delta(x,x')\right.\nonumber
\\
& &\left.\qquad+\frac{D^2}{4} [\Box_x\Delta(x,x')]^2\right]\,,
\label{AdlerZee}
\end{eqnarray}
where $w=x-x'$.
The equations (\ref{cosc}) and (\ref{AdlerZee}) are our main results in this paper.

The Green's function of a canonical free massive scalar field can be written by
an integral form with the so-called Schwinger parameter
\begin{equation}
\Delta_k(x,x')=\int_0^\infty\frac{ds}{(4\pi
s)^{D/2}}\exp\left[%\textstyle
-\frac{|w|^2}{4s}-m_k^2s\right]
\,,
\end{equation}
where the symmetry $\Delta(x,x')=\Delta(x',x)$ is apparent.
Using this form, we can simplify the formulas (\ref{cosc}) and (\ref{AdlerZee}).
The details are shown in Appendix \ref{cig}.

Performing straightforward differentiations (see Appendix \ref{cig}) and taking
a naive limitation
$x'\rightarrow x$ i.e.,
$|w|\rightarrow 0$, we obtain
\begin{equation}
\frac{\Lambda_{ind}}{8\pi G_{ind}}=-
\int_0^\infty\frac{ds}{2(4\pi
)^{D/2}s^{D/2+1}}\varrho(s)
\,,\label{i1}
\end{equation}
where
\begin{equation}
\varrho(s)\equiv\sum_{k=1}^n e^{-m_k^2s}
\,,
\end{equation}
is a useful abbreviation.

The calculation of the induced gravitational constant is obtained after lengthy but
straightforward calculations (see Appendix \ref{cig}).
The result is
\begin{equation}
\frac{1}{16\pi G_{ind}}=\frac{1}{12(4\pi)^{D/2}}\int_0^\infty\frac{ds}{s^{D/2}}
\varrho(s)
\,.\label{i2}
\end{equation}

These results confirm the 
equivalence of the heat kernel method and the Adler--Zee formula previously used in
the literature
\cite{Adler1,Adler2,Adler3,Zee1,Zee2,MW,DS,FF,Visser,BS,KS1,KS2,KKS,DM}.

The integral forms (\ref{i1}) and (\ref{i2}) are compact but include
divergences as is known in the literature.
The finite part of both the cosmological constant and the gravitational constant
has often been reported in the model of the Kaluza--Klein theory and its
generalization \cite{Toms,CW}, as known calculable examples.
In practice, we should manage to cancel the divergent part by introducing
massive scalar, fermion, and vector fields \cite{FF,KS2,KKS,KPV}
and even other compensating fields \cite{KPV}.
In this paper, though we focused on the scalar field theory,
the higher-derivative generalization of spinor or vector field theories is
feasible \textit{\`a la} the generalized Lee--Wick theory \cite{GOW,CL,Carone}.
The concrete example of cancellation of divergences is not attempted in
this study.

In the next section in our present paper, we examine the induced gravity from
certain scalar models with an infinite number of derivatives using the expression
obtained above, and show that the induced cosmological and gravitational constants
are calculable in the models.

%%%%%%%%%%%%%%%%%%%%%%%%%%%%%%%%%%%%%%%%%%%%%%%%%%%%%%%%%%%%%%%%%%%%%%%%%%%
%%%%%%%%%%%%%%%%%%%%%%%%%%%%%%%%%%%%%%%%%%%%%%%%%%%%%%%%%%%%%%%%%%%%%%%%%%%
%%%%%%%%%%%%%%%%%%%%%%%%%%%%%%%%%%%%%%%%%%%%%%%%%%%%%%%%%%%%%%%%%%%%%%%%%%%
\section{induced cosmological and gravitational constants from an infinite
derivative scalar field theory}
\label{sec3}
%%%%%%%%%%%%%%%%%%%%%%%%%%%%%%%%%%%%%%%%%%%%%%%%%%%%%%%%%%%%%%%%%%%%%%%%%%%
%%%%%%%%%%%%%%%%%%%%%%%%%%%%%%%%%%%%%%%%%%%%%%%%%%%%%%%%%%%%%%%%%%%%%%%%%%%
%%%%%%%%%%%%%%%%%%%%%%%%%%%%%%%%%%%%%%%%%%%%%%%%%%%%%%%%%%%%%%%%%%%%%%%%%%%

In this section, we introduce certain examples of higher-derivative theory.
We first consider the following Lagrangian with higher derivatives
on a real scalar field $\phi$:
\begin{equation}
\mathcal{L}=-\frac{1}{2}\phi(x)\frac{\sqrt{-\Box}\sinh[\pi l\sqrt{-\Box}]}{{\pi}l}
\phi(x)=-\frac{1}{2}\phi(x)\frac{1}{l^2|\Gamma(il\sqrt{-\Box})|^2}\phi(x)\,.
\label{ThisL}
\end{equation}
Although this concise form of the Lagrangian seems to include
the square root of the differential operator, there is actually no singular
operator, as seen from the series expansion.
The Green's function in momentum space of this
model first appeared in Refs.~\cite{FS1,FS2,Fujikawa}. 
This falls in a type of the Pauli--Villars regularized function, since
a formula of an alternating series shows
\begin{equation}
\frac{\pi l}{\sqrt{-\Box}\sinh[\pi l\sqrt{-\Box}]}=\frac{1}{-\Box}+
2\sum_{k=1}^{\infty}\frac{(-1)^k}{-\Box+\frac{k^2}{l^2}}\,.
\end{equation}
Therefore, we can expect better UV behavior of physical quantities connected to
this function, so it is suitable as the first model to be studied.
 Incidentally,
the Green's function in flat configuration space can be written as \cite{KKSW1}
\begin{equation}
G(x,x')=\frac{l}{\pi^{\frac{D-1}{2}}}\sum_{k=0}^\infty
\frac{\Gamma\left(\frac{D-1}{2}\right)}{\left[|x-x'|^2+4\pi^2l^2\left(k+\frac{1}{2}
\right)^2\right]^{\frac{D-1}{2}}}\,.
\end{equation}
It can be confirmed that the limit $l\rightarrow 0$ reduces this expression to the
canonical massless Green's function.

The Lagrangian (\ref{ThisL}) can be written by using the infinite product as
\begin{equation}
\mathcal{L}=-\frac{1}{2C}\phi(x)(-\Box)\prod_{k=1}^{\infty}\left[
-\Box+\frac{k^2}{l^2}\right]\phi(x)\,,
\end{equation}
where $C=\prod_{k=1}^\infty\frac{k^2}{l^2}.$
Further, in order to apply the treatment according to Ref.~\cite{GPS} and the
formulation in the previous section, we rewrite this in the form
\begin{equation}
\mathcal{L}=-\frac{1}{2C}\phi(x)\left[\prod_{i=1}^{\infty}A_i\right]\phi(x)\,,
\end{equation}
where $A_i=-\Box+m_i^2$, $m_i^2=(i-1)^2/l^2~(i=1, 2, 3, \ldots)$. Note that $m_1=0$
in the present model.

A primitive calculation with the formula (\ref{i1}) gives, after the integration
on $s$,
\begin{eqnarray}
\frac{\Lambda_{ind}}{8\pi G_{ind}}&=&-\frac{\Gamma(-D/2)}{2(4\pi
)^{D/2}}\sum_{k=1}^\infty m_k^{D}\nonumber \\
&=&-\frac{\Gamma(-D/2)}{2(4\pi
)^{D/2}l^D}\sum_{k=1}^\infty k^{D}=-\frac{\Gamma(-D/2)\zeta_R(-D)}{2(4\pi
)^{D/2}l^D}\nonumber \\
&=&-\frac{\Gamma((D+1)/2)\zeta_R(D+1)}{2\pi^{D+1/2}(4\pi
)^{D/2}l^D}\,,\label{zeta1}
\end{eqnarray}
where $\zeta_R(z)$ is the Riemann's zeta function.
In the last equality, we used the mathematical formula
$\zeta_R(z)\Gamma(z/2)=\pi^{z-1/2}\zeta(1-z)\Gamma((1-z)/2)$.
In this way, the infinite sum can often be evaluated as a finite value,
as in the Kaluza--Klein theories \cite{Toms,CW}, despite there being a divergence. 

In this case, it is possible to consider the sum in $\varrho(s)$ first.
Then, the mathematical formula leads to 
\begin{equation}
\varrho(s)\equiv\sum_{k=1}^\infty
e^{-m_k^2s}=\frac{1}{2}\left[1+\vartheta_3(0,e^{-s/l^2})\right]
=\frac{1}{2}\left[1+l\sqrt{\frac{\pi}{s}}\vartheta_3(0,e^{-\pi^2l^2/s})\right]
\,,
\end{equation}
where $\vartheta_a(v,q)$ is the Jacobi theta function.
If we naively (but as in the zeta function regularization \cite{Hawking}) discard
the divergent contribution, we get
\begin{eqnarray}
& &\frac{\Lambda_{ind}}{8\pi G_{ind}}=-
\int_0^\infty\frac{ds}{2(4\pi
)^{D/2}s^{D/2+1}}\varrho(s)\nonumber \\
& &\Rightarrow
-\int_0^\infty\frac{l\sqrt{\pi}ds}{2(4\pi
)^{D/2}s^{(D+1)/2+1}}\sum_{k=1}^\infty\exp\left(-\frac{\pi^2l^2k^2}{s}\right)
=-\frac{\Gamma((D+1)/2)\zeta_R(D+1)}{2\pi^{D+1/2}(4\pi
)^{D/2}l^D}
\,,
\end{eqnarray}
which reproduces exactly same result as (\ref{zeta1}).
For $D=4$, the numerical value is found to be $\frac{\Lambda_{ind}}{8\pi
G_{ind}}=-2.5279
\times 10^{-5} l^{-4}$.

%%%%%%%%%%%%%%%%%%%%%%%%%%%%%%%%%%%%%%%%%%%%%%%%%%%%%%%%%%%%%%%%%%%%%%%%%%%
%%%%%%%%%%%%%%%%%%%%%%%%%%%%%%%%%%%%%%%%%%%%%%%%%%%%%%%%%%%%%%%%%%%%%%%%%%%

Similarly, the primitive calculation of the induced gravitational constant gives
\begin{eqnarray}
\frac{1}{16\pi G_{ind}}&=&\frac{\Gamma(1-D/2)}{12(4\pi
)^{D/2}}\sum_{k=1}^\infty m_k^{D-2}\nonumber \\
&=&\frac{\Gamma(1-D/2)}{12(4\pi
)^{D/2}l^{D-2}}\sum_{k=1}^\infty k^{D-2}=\frac{\Gamma(1-D/2)\zeta_R(2-D)}{12(4\pi
)^{D/2}l^{D-2}}\nonumber \\
&=&\frac{\Gamma((D-1)/2)\zeta_R(D-1)}{12\pi^{D-3/2}(4\pi
)^{D/2}l^{D-2}}\,.
\end{eqnarray}
This result is also obtained by discarding the divergent contribution in
$\varrho(s)$ in (\ref{i2}), as
\begin{eqnarray}
& &\frac{1}{16\pi
G_{ind}}=\frac{1}{12(4\pi)^{D/2}}\int_0^\infty\frac{ds}{s^{D/2}}\varrho(s)\nonumber
\\ & &\Rightarrow \int_0^\infty\frac{l\sqrt{\pi}ds}{12(4\pi
)^{D/2}s^{(D+1)/2}}\sum_{k=1}^\infty\exp\left(-\frac{\pi^2l^2k^2}{s}\right)
=\frac{\Gamma((D-1)/2)\zeta_R(D-1)}{12\pi^{D-3/2}(4\pi
)^{D/2}l^{D-2}}
\,.
\end{eqnarray}

For $D=4$, the numerical value is $\frac{1}{16\pi
G_{ind}}=+3.21362\times 10^{-5} l^{-2}$.

%%%%%%%%%%%%%%%%%%%%%%%%%%%%%%%%%%%%%%%%%%%%%%%%%%%%%%%%%%%%%%%%%%%%%%%%%%%
%%%%%%%%%%%%%%%%%%%%%%%%%%%%%%%%%%%%%%%%%%%%%%%%%%%%%%%%%%%%%%%%%%%%%%%%%%%

Another model with an infinite number of derivatives is
\begin{equation}
\mathcal{L}=-\frac{1}{2}\phi(x){\cosh[\pi l\sqrt{-\Box}]}
\phi(x)=-\frac{1}{2}\phi(x)\frac{\pi}{|\Gamma(\frac{1}{2}+il\sqrt{-\Box})|^2}\phi(x)\,.
\end{equation}
Note that this model does not have a massless mode.
The Lagrangian of this model can be written by using the infinite product as
\begin{equation}
\mathcal{L}=-\frac{1}{2C'}\phi(x)\prod_{k=1}^{\infty}\left[
-\Box+\frac{(k-1/2)^2}{l^2}\right]\phi(x)\,,
\end{equation}
where 
$C'=\prod_{k=1}^\infty\frac{(k-1/2)^2}{l^2}$.
Then, the mass spectrum is given by $m_i^2=(i-1/2)^2/l^2~(i=1, 2, 3, \ldots)$
and we find 
\begin{equation}
\varrho(s)\equiv\sum_{k=1}^\infty
e^{-m_k^2s}=\frac{1}{2}\vartheta_2(0,e^{-s/l^2})
=\frac{1}{2}l\sqrt{\frac{\pi}{s}}\vartheta_4(0,e^{-\pi^2l^2/s})
\,,
\end{equation}
where $\vartheta_a(v,q)$ is the Jacobi theta function.

Therefore, the finite part of the induced cosmological constant in this case can be
obtained by
\begin{eqnarray}
\frac{\Lambda_{ind}}{8\pi G_{ind}}&=&-\frac{\Gamma(-D/2)}{2(4\pi
)^{D/2}}\sum_{k=1}^\infty m_k^{D}\nonumber \\
&=&-\frac{\Gamma(-D/2)}{2(4\pi
)^{D/2}l^D}\sum_{k=1}^\infty
\left(k-\frac{1}{2}\right)^{D}=\left(1-\frac{1}{2^D}\right)\frac{\Gamma(-D/2)\zeta_R(-D)}{2(4\pi
)^{D/2}l^D}\nonumber
\\
&=&\left(1-\frac{1}{2^D}\right)\frac{\Gamma((D+1)/2)\zeta_R(D+1)}{2\pi^{D+1/2}(4\pi
)^{D/2}l^D}\,,
\end{eqnarray}
or
\begin{eqnarray}
\frac{\Lambda_{ind}}{8\pi G_{ind}}&=&-
\int_0^\infty\frac{ds}{2(4\pi
)^{D/2}s^{D/2+1}}\varrho(s)\nonumber \\
&\Rightarrow&
-\int_0^\infty\frac{l\sqrt{\pi}ds}{2(4\pi
)^{D/2}s^{(D+1)/2+1}}\sum_{k=1}^\infty(-1)^{k}\exp\left(-\frac{\pi^2l^2k^2}{s}
\right)\nonumber \\
&=&\left(1-\frac{1}{2^D}\right)\frac{\Gamma((D+1)/2)\zeta_R(D+1)}{2\pi^{D+1/2}(4\pi
)^{D/2}l^D}
\,.
\end{eqnarray}
For $D=4$, the numerical value is $\frac{\Lambda_{ind}}{8\pi
G_{ind}}=+2.36991\times 10^{-5} l^{-4}$.

Similarly, the finite part of the induced gravitational constant can be obtained by
\begin{eqnarray}
\frac{1}{16\pi G_{ind}}&=&\frac{\Gamma(1-D/2)}{12(4\pi
)^{D/2}}\sum_{k=1}^\infty m_k^{D-2}\nonumber \\
&=&\frac{\Gamma(1-D/2)}{12(4\pi
)^{D/2}l^{D-2}}\sum_{k=1}^\infty
\left(k-\frac{1}{2}\right)^{D-2}=-\left(1-\frac{1}{2^{D-2}}\right)\frac{\Gamma(1-D/2)\zeta_R(2-D)}{12(4\pi
)^{D/2}l^{D-2}}\nonumber
\\
&=&-\left(1-\frac{1}{2^{D-2}}\right)\frac{\Gamma((D-1)/2)\zeta_R(D-1)}{12\pi^{D-3/2}(4\pi
)^{D/2}l^{D-2}}\,,
\end{eqnarray}
or
\begin{eqnarray}
\frac{1}{16\pi
G_{ind}}&=&\frac{1}{12(4\pi)^{D/2}}\int_0^\infty\frac{ds}{s^{D/2}}\varrho(s)\nonumber
\\ &=& \int_0^\infty\frac{l\sqrt{\pi}ds}{12(4\pi
)^{D/2}s^{(D+1)/2}}\sum_{k=1}^\infty(-1)^k\exp\left(-\frac{\pi^2l^2k^2}{s}\right)\nonumber
\\ 
&=&-\left(1-\frac{1}{2^{D-2}}\right)\frac{\Gamma((D-1)/2)\zeta_R(D-1)}{12\pi^{D-3/2}(4\pi
)^{D/2}l^{D-2}}
\,.
\end{eqnarray}
For $D=4$, its numerical value is $\frac{1}{16\pi
G_{ind}}=-2.41021\times 10^{-5} l^{-2}$.

The further study on other models which have various mass spectra and extension to
the spinor and vector field theories is left for a future work.
Please note that the sum of contributions from various matter fields 
determines the induced constants.%
\footnote{Very powerful methods for calculating effective actions, including
background gravitational fields, can be found in Ref.~\cite{BSbook}.}

%%%%%%%%%%%%%%%%%%%%%%%%%%%%%%%%%%%%%%%%%%%%%%%%%%%%%%%%%%%%%%%%%%%%%%%%%%%
%%%%%%%%%%%%%%%%%%%%%%%%%%%%%%%%%%%%%%%%%%%%%%%%%%%%%%%%%%%%%%%%%%%%%%%%%%%
%%%%%%%%%%%%%%%%%%%%%%%%%%%%%%%%%%%%%%%%%%%%%%%%%%%%%%%%%%%%%%%%%%%%%%%%%%%
\section{Conclusion}
\label{conclusion}
%%%%%%%%%%%%%%%%%%%%%%%%%%%%%%%%%%%%%%%%%%%%%%%%%%%%%%%%%%%%%%%%%%%%%%%%%%%
%%%%%%%%%%%%%%%%%%%%%%%%%%%%%%%%%%%%%%%%%%%%%%%%%%%%%%%%%%%%%%%%%%%%%%%%%%%
%%%%%%%%%%%%%%%%%%%%%%%%%%%%%%%%%%%%%%%%%%%%%%%%%%%%%%%%%%%%%%%%%%%%%%%%%%%

In this paper, we have studied the theory of induced gravity derived from the
higher-derivative theory. In the higher-derivative scalar field theory
without self-interaction, it is concluded that the induced
cosmological and gravitational constants are calculated as a sum of those obtained
from free scalar fields, whose number corresponds to the number of d'Alembertian
acting on the scalar field. In addition, we have confirmed that the Adler--Zee
formula and the heat kernel method give the same result in arbitrary dimensions,
and shown that a finite value for the induced constants can be extracted in the
model with infinite derivatives.
Unlike the Kaluza--Klein theory, we can consider models with arbitrary masses $m_k$
in the higher-derivative theories\footnote{That is the fundamental scale $l$ used
in our models should not be the Planck length.} 
and we can even use copies of multiple models which result in different signs and
magnitudes of induced quantities.  
%For this reason, it is interesting to
%compare various induced gravity models with observation. 
The higher-derivative
theories can be found in modified gravity and string theory, though the structures
of them are more complicated than the models considered in this paper. We would
like to take this work as the first step in exploring quantum effects in
more general models of the higher-derivative theory.
 
As a future issue, first of all, we will consider the introduction of fields with
other spins, similar to the conventional induced gravity model including
Ref.~\cite{KPV}, for other fields can cancel divergences. 
Our investigations so far
have only been on free higher-derivative theories.
In addition, since the higher-derivative theory with self-interaction has a
complicated structure even considering an effective theory, the calculation of the
quantum effect becomes generally non-trivial. One may need to think about a theory
with some useful symmetry. By the way, in general, in the theory of
higher-derivative fields coupled to the curvature, the calculation of the induced
gravitational constant becomes quite nontrivial.%
\footnote{The induced gravity from the superrenormalizable models
\cite{Modesto1,BT,ALS,MS,Modesto2} has not been considered by anyone and this
could also be decent work to do.} We would like to examine this case by
constructing various models.
 
Extensions in different directions include mathematical applications to recent
discussions of entropic field theory \cite{CMT,CTT1,CTT2}, induced gravity effect
in continuous mass distribution theory \cite{Georgi1,Georgi2,Krasnikov}, and so on.
These topics are also reserved for future work.

%%%%%%%%%%%%%%%%%%%%%%%%%%%%%%%%%%%%%%%%%%%%%%%%%%%%%%%%%%%%%%%%%%%%%%%%%%%
%%%%%%%%%%%%%%%%%%%%%%%%%%%%%%%%%%%%%%%%%%%%%%%%%%%%%%%%%%%%%%%%%%%%%%%%%%%

%%%%%%%%%%%%%%%%%%%%%%%%%%%%%%%%%%%%%%%%%%%%%%%%%%%%%%%%%%%%%%%%%
%%%%%%%%%%%%%%%%%%%%%%%%%%%%%%%%%%%%%%%%%%%%%%%%%%%%%%%%%%%%%%%%%
\appendix
%%%%%%%%%%%%%%%%%%%%%%%%%%%%%%%%%%%%%%%%%%%%%%%%%%%%%%%%%%%%%%%%%

%%%%%%%%%%%%%%%%%%%%%%%%%%%%%%%%%%%%%%%%%%%%%%%%%%%%%%%%%%%%%%%%%%%%%%%%%%%
%%%%%%%%%%%%%%%%%%%%%%%%%%%%%%%%%%%%%%%%%%%%%%%%%%%%%%%%%%%%%%%%%%%%%%%%%%%
%%%%%%%%%%%%%%%%%%%%%%%%%%%%%%%%%%%%%%%%%%%%%%%%%%%%%%%%%%%%%%%%%%%%%%%%%%%
\section{The derivation of (\ref{AZ1}) and (\ref{AZ2})}
\label{AZR}
We briefly review the derivation of the Adler--Zee formulas (\ref{AZ1}) and
(\ref{AZ2}) in this Appendix (for a review, see Refs.~\cite{MW,DM}). 

In the weak-field limit, the action is expanded as
\begin{eqnarray}
S[\phi,g_{\mu\nu}]&=&S[\phi,\eta_{\mu\nu}]+\int d^Dx\left.\frac{\delta S}{\delta
g_{\mu\nu}(x)}\right|_{\phi,\eta_{\alpha\beta}}h_{\mu\nu}(x)\nonumber\\
\qquad& &
+\frac{1}{2}\int d^Dx\int d^Dy\left.\frac{\delta^2 S}{\delta
g_{\mu\nu}(x)\delta
g_{\rho\sigma}(y)}\right|_{\phi,\eta_{\alpha\beta}}h_{\mu\nu}(x)h_{\rho\sigma}(y)+\cdots\,,
\end{eqnarray}
where $g_{\mu\nu}=\eta_{\mu\nu}+h_{\mu\nu}$.
Then, the effective action $S[h]$ is given by
\begin{eqnarray}
S[h]&=&-\ln\int[D\phi]e^{-S[\phi,g_{\mu\nu}]}\nonumber \\
&=&\frac{1}{2}\int d^Dx\, h^{\mu\nu}(x)\langle T_{\mu\nu}(x)\rangle
+\frac{1}{4}\int d^Dx h^{\mu\nu}(x)h_{\rho\sigma}(x)\langle
\frac{\delta T_{\mu\nu}(x)}{\delta g_{\rho\sigma}(y)}\rangle\nonumber \\
& &-\frac{1}{8}\int d^Dx\int d^Dy\, h^{\mu\nu}(x)h^{\rho\sigma}(y)
\langle \bar{T}_{\mu\nu}(x)\bar{T}_{\rho\sigma}(y)\rangle+\cdots\,,
\end{eqnarray}
where we used the stress tensor $T^{\mu\nu}(x)=\frac{2}{\sqrt{|g|}}\frac{\delta
S}{\delta g_{\mu\nu}(x)}$.  $\langle \cdots\rangle$ denotes a vacuum expectation
value and
$\bar{T}_{\mu\nu}\equiv T_{\mu\nu}-\langle T_{\mu\nu}\rangle$.
Note that we are working with $D$-dimensional
Euclidean space ($\eta_{\mu\nu}=\delta_{\mu\nu}$).
For simplicity, we specialize the metric as
$h_{\mu\nu}(x)=\frac{1}{D}\eta_{\mu\nu}h(x)$ (Refs.~\cite{Zee1,MW,DM}) and then 
we can express the effective action as a functional of $h(x)$.

The part of the effective action which includes no derivative of $h$ is
given by
\begin{equation}
\int d^Dx \frac{1}{2D}h(x)\langle T(x)\rangle+O(h^2)\,,
\end{equation}
while the part of the effective action which includes two derivatives of $h$ is
\begin{equation}
-\int d^Dx \frac{1}{16D^3}h(x)\Box h(x)\int d^Dw |w|^2\langle
\bar{T}(x)\bar{T}(y)\rangle+\cdots\,,
\end{equation}
where $T\equiv \eta^{\mu\nu}T_{\mu\nu}$.
Here we used the Taylor expansion $h(y)=h(x)-w^\mu\partial_\mu
h(x)+\frac{1}{2}w^\mu w^\nu\partial_\mu\partial_\nu h(x)$, where
$w=x-y$, and the isotropy of the space, i.e., $w^\mu w^\nu\rightarrow
\eta^{\mu\nu}|w|^2/D$ in the integral.

The comparison of these expression with
\begin{equation}
\int d^Dx \sqrt{|g|}=\int d^Dx \left(1+\frac{1}{2}h(x)+O(h^2)\right)\,,
\end{equation}
and
\begin{equation}
\int d^Dx \sqrt{|g|}R=\frac{(D-1)(D-2)}{4D^2}\int d^Dx\,
h(x)\Box h(x)+\cdots\,,
\end{equation}
(where $R$ is the scalar curvature) gives (\ref{AZ1}) and (\ref{AZ2}).

%%%%%%%%%%%%%%%%%%%%%%%%%%%%%%%%%%%%%%%%%%%%%%%%%%%%%%%%%%%%%%%%%%%%%%%%%%%
%%%%%%%%%%%%%%%%%%%%%%%%%%%%%%%%%%%%%%%%%%%%%%%%%%%%%%%%%%%%%%%%%%%%%%%%%%%
%%%%%%%%%%%%%%%%%%%%%%%%%%%%%%%%%%%%%%%%%%%%%%%%%%%%%%%%%%%%%%%%%%%%%%%%%%%

%%%%%%%%%%%%%%%%%%%%%%%%%%%%%%%%%%%%%%%%%%%%%%%%%%%%%%%%%%%%%%%%%%%%%%%%%%%
%%%%%%%%%%%%%%%%%%%%%%%%%%%%%%%%%%%%%%%%%%%%%%%%%%%%%%%%%%%%%%%%%%%%%%%%%%%
%%%%%%%%%%%%%%%%%%%%%%%%%%%%%%%%%%%%%%%%%%%%%%%%%%%%%%%%%%%%%%%%%%%%%%%%%%%
\section{The stress tensor in the $n=2$ Lee--Wick scalar field theory}
\label{LeeW}
%%%%%%%%%%%%%%%%%%%%%%%%%%%%%%%%%%%%%%%%%%%%%%%%%%%%%%%%%%%%%%%%%%%%%%%%%%%
%%%%%%%%%%%%%%%%%%%%%%%%%%%%%%%%%%%%%%%%%%%%%%%%%%%%%%%%%%%%%%%%%%%%%%%%%%%
%%%%%%%%%%%%%%%%%%%%%%%%%%%%%%%%%%%%%%%%%%%%%%%%%%%%%%%%%%%%%%%%%%%%%%%%%%%

We can introduce an auxiliary field to study the $n=2$ Lee--Wick scalar field
theory, referring works on higher-derivative theories \cite{GOW} motivated by
the seminal Lee--Wick model \cite{LW1,LW2,LW3}.
We assume the Lagrangian
\begin{equation}
{\cal
L}=-\frac{1}{2(m_2^2-m_1^2)}\phi(-\Box+m_1^2)(-\Box+m_2^2)\phi
\,.
\end{equation}
We also consider the alternative Lagrangian
\begin{equation}
{\cal
L}'=-\frac{1}{2}\phi(-\Box+m_1^2)\phi-
\psi(-\Box+m_2^2)\phi+\frac{1}{2}(m_2^2-m_1^2)\psi^2
\,.\label{al}
\end{equation}
The equation of motion of the initially auxiliary field $\psi$ from the Lagrangian
(\ref{al}) is
\begin{equation}
\psi=\frac{1}{m_2^2-m_1^2}(-\Box+m_1^2)\phi
\,.\label{p1}
\end{equation}
Thus, substituting this equation to the alternative Lagrangian ${\cal L}'$ yields
the original Lagrangian
${\cal L}$. On the other hand, defining the field $\chi$ as
\begin{equation}
\chi=\phi+\psi=\frac{1}{m_2^2-m_1^2}(-\Box+m_2^2)\phi
\,,\label{c1}
\end{equation}
the Lagrangian (\ref{al}) is also written by
\begin{equation}
{\cal
L}'=-\frac{1}{2}\chi(-\Box+m_1^2)\chi+\frac{1}{2}\psi(-\Box+m_2^2)\psi
\,.\label{ll}
\end{equation}
This effective Lagrangian describes two free scalar fields, whose masses are
$m_1$ and $m_2$. The kinetic term for the scalar field $\psi$ with mass
$m_2$ has a ``wrong'' sign.
This field decouples from the physical spectrum if $m_2\rightarrow\infty$, since
its mass becomes infinitely large. 

The stress tensor derived from the last Lagrangian (\ref{ll}) reads
\begin{equation}
T_{\mu\nu}=\partial_\mu\chi\partial_\nu\chi-\frac{1}{2}g_{\mu\nu}
(\partial_\rho\chi\partial^\rho\chi+m_1^2\chi^2)-
\partial_\mu\psi\partial_\nu\psi+\frac{1}{2}g_{\mu\nu}
(\partial_\rho\psi\partial^\rho\psi+m_2^2\psi^2)\,.
\end{equation}
By substituting (\ref{p1}) and (\ref{c1}) into this,
we obtain
\begin{eqnarray}
T_{\mu\nu}&=&\frac{1}{m_2^2-m_1^2}\Bigl[
-\partial_\mu\phi\partial_\nu\Box\phi-\partial_\nu\phi\partial_\mu\Box\phi
+g_{\mu\nu}\partial_\rho\phi\partial^\rho\Box\phi
+{\textstyle\frac{1}{2}}g_{\mu\nu}(\Box\phi)^2\nonumber \\
&
&\qquad\qquad+(m_1^2+m_2^2)\Bigl(\partial_\mu\phi\partial_\nu\phi-{\textstyle\frac{1}{2}}
g_{\mu\nu}
\partial_\rho\phi\partial^\rho\phi\Bigr)-{\textstyle\frac{1}{2}}
g_{\mu\nu}
m_1^2m_2^2\phi^2\Bigr]\,,
\end{eqnarray}
which coincides with the result found in Refs.~\cite{CK,KLS,GPS}.
Incidentally, the trace of the stress tensor becomes
\begin{eqnarray}
T&=&\frac{1}{m_2^2-m_1^2}\Bigl[
(D-2)\partial_\rho\phi\partial^\rho\Box\phi
-(m_1^2+m_2^2)
{\textstyle\frac{D-2}{2}}
\partial_\rho\phi\partial^\rho\phi+{\textstyle\frac{D}{2}}(\Box\phi)^2-{\textstyle\frac{D}{2}}
m_1^2m_2^2\phi^2\Bigr]\nonumber \\
&=&\frac{1}{m_2^2-m_1^2}\Bigl\{
-\frac{D-2}{2}\Bigl[\partial_\rho\phi\partial^\rho(-\Box+m_1^2)\phi
+\partial_\rho\phi\partial^\rho(-\Box+m_2^2)\phi\Bigr]
\nonumber \\
&
&\quad\quad-{\frac{D}{2}}\Bigl[(-\Box+m_1^2)\phi\Box\phi
+\phi\Box(-\Box+m_2^2)\phi
+
\phi(-\Box+m_1^2)(-\Box+m_2^2)\phi\Bigr]\Bigr\}\,,
\end{eqnarray}
which agrees with (\ref{TT}) for $n=2$ up to the last term that can be discarded by
the field equation.

%%%%%%%%%%%%%%%%%%%%%%%%%%%%%%%%%%%%%%%%%%%%%%%%%%%%%%%%%%%%%%%%%%%%%%%%%%%

%%%%%%%%%%%%%%%%%%%%%%%%%%%%%%%%%%%%%%%%%%%%%%%%%%%%%%%%%%%%%%%%%%%%%%%%%%%
%%%%%%%%%%%%%%%%%%%%%%%%%%%%%%%%%%%%%%%%%%%%%%%%%%%%%%%%%%%%%%%%%%%%%%%%%%%
%%%%%%%%%%%%%%%%%%%%%%%%%%%%%%%%%%%%%%%%%%%%%%%%%%%%%%%%%%%%%%%%%%%%%%%%%%%
\section{The proof of (\ref{AdlerZee})}
\label{prf}
%%%%%%%%%%%%%%%%%%%%%%%%%%%%%%%%%%%%%%%%%%%%%%%%%%%%%%%%%%%%%%%%%%%%%%%%%%%
%%%%%%%%%%%%%%%%%%%%%%%%%%%%%%%%%%%%%%%%%%%%%%%%%%%%%%%%%%%%%%%%%%%%%%%%%%%
%%%%%%%%%%%%%%%%%%%%%%%%%%%%%%%%%%%%%%%%%%%%%%%%%%%%%%%%%%%%%%%%%%%%%%%%%%%

By rewriting (\ref{TT2}), we find
\begin{equation}
\bar{T}(x)=\lim_{x'\rightarrow x}\sum_{k=0}^n
C^{-1}\mathcal{O}_{xx'}\mbox{:}\eta_k(x)
\chi_{k+1}(x')\mbox{:}\,,
\end{equation}
where%
\footnote{As previously mentioned in the text, $\Box_x$ can read
$(\Box_x+\Box_{x'})/2$.}
\begin{equation}
\mathcal{O}_{xx'}=-\frac{D-2}{2}
\partial_{\rho}\partial^{\rho'}-\frac{D}{2}\Box_x\,,
\end{equation}
and $\mbox{:}~\mbox{:}$ stands for a normal ordered product.
Therefore, we would like to know 
$\sum_{k=0}^n\sum_{l=0}^n\langle $ $\mbox{:}\eta_k(x) 
\chi_{k+1}(x')\mbox{:}$ $\mbox{:}\eta_l(y)
\chi_{l+1}(y')\mbox{:}$ $\rangle$ for an evaluation from the Adler--Zee formula. 

To this end, we start with
\begin{eqnarray}
& &\langle \mbox{:}\eta_0(x)\chi_{1}(x')+\eta_1(x)\chi_{2}(x')\mbox{:}\,
\mbox{:}\eta_0(y)\chi_{1}(y')+\eta_1(y)\chi_{2}(y')\mbox{:}\rangle\nonumber \\
&=&\langle\eta_0(x)\eta_0(y)\rangle
\langle\chi_{1}(x')\chi_{1}(y')\rangle+\langle\eta_0(x)\chi_{1}(y')\rangle
\langle\eta_0(y)\chi_{1}(x')\rangle\nonumber \\
&+&\langle\eta_1(x)\eta_1(y)\rangle
\langle\chi_{2}(x')\chi_{2}(y')\rangle+\langle\eta_1(x)\chi_{2}(y')\rangle
\langle\eta_1(y)\chi_{2}(x')\rangle\nonumber \\
&+&2\langle\eta_0(x)\eta_1(y)\rangle
\langle\chi_{1}(x')\chi_{2}(y')\rangle+2\langle\eta_0(x)\chi_{2}(y')\rangle
\langle\eta_1(y)\chi_{1}(x')\rangle%\nonumber \\
%&+&\langle\eta_1(x)\eta_0(y)\rangle
%\langle\chi_{2}(x')\chi_{1}(y')\rangle+\langle\eta_1(x)\chi_{1}(y')\rangle
%\langle\eta_0(y)\chi_{2}(x')\rangle
\,,\label{gt}
\end{eqnarray}
where we used the symmetry under $(x,x')\leftrightarrow(y,y')$.
Now, we use $\eta_0=\phi$, $\eta_1=A_1\phi$, $\chi_1=A_2\cdots A_n\phi$, 
$\chi_2=A_3\cdots A_n\phi$, and $\langle\phi(x)\phi(x')\rangle=\frac{C}{A_1\cdots
A_n}\mathbf{1}_{xx'}$, in the symbolic notation.
Then, we find
\begin{eqnarray}
& &C^{-2}\langle\eta_0(x)\eta_0(y)\rangle
\langle\chi_{1}(x')\chi_{1}(y')\rangle=\frac{1}{A_1\cdots A_n}\mathbf{1}_{xy}
\frac{1}{A_1}{A_2\cdots A_n}\mathbf{1}_{x'y'}\label{e1}\\
& &C^{-2}\langle\eta_0(x)\chi_{1}(y')\rangle
\langle\eta_0(y)\chi_{1}(x')\rangle=\frac{1}{A_1}\mathbf{1}_{xy'}
\frac{1}{A_1}\mathbf{1}_{x'y}\label{e2}\\ 
& &C^{-2}\langle\eta_1(x)\eta_1(y)\rangle
\langle\chi_{2}(x')\chi_{2}(y')\rangle=A_1\frac{1}{A_2\cdots A_n}\mathbf{1}_{xy}
\frac{1}{A_1A_2}{A_3\cdots A_n}\mathbf{1}_{x'y'}\label{e3}\\
& &C^{-2}\langle\eta_1(x)\chi_{2}(y')\rangle
\langle\eta_1(y)\chi_{2}(x')\rangle=\frac{1}{A_2}\mathbf{1}_{xy'}
\frac{1}{A_2}\mathbf{1}_{x'y}\label{e4} \\
& &C^{-2}\langle\eta_0(x)\eta_1(y)\rangle
\langle\chi_{1}(x')\chi_{2}(y')\rangle=\frac{1}{A_2\cdots A_n}\mathbf{1}_{xy}
\frac{1}{A_1}{A_3\cdots A_n}\mathbf{1}_{x'y'}\label{e5}\\
& &C^{-2}\langle\eta_0(x)\chi_{2}(y')\rangle
\langle\eta_1(y)\chi_{1}(x')\rangle=\frac{1}{A_1A_2}\mathbf{1}_{xy'}
\mathbf{1}_{x'y}\label{e6}
\,.
\end{eqnarray}

Further, after using $\frac{A_2}{A_1}=1+\frac{m_2^2-m_1^2}{A_1}$ in (\ref{e1}),
 $\frac{A_1}{A_2}=1-\frac{m_2^2-m_1^2}{A_2}$ in (\ref{e3}),
and 
$\frac{1}{A_1A_2}=\frac{1}{m_2^2-m_1^2}\left(\frac{1}{A_1}-\frac{1}{A_2}\right)$
in (\ref{e1}) and (\ref{e3}), the result for (\ref{gt}) turns out to be
\begin{eqnarray}
& &C^{-2}\langle \mbox{:}\eta_0(x)\chi_{1}(x')+\eta_1(x)\chi_{2}(x')\mbox{:}
\mbox{:}\eta_0(y)\chi_{1}(y')+\eta_1(y)\chi_{2}(y')\mbox{:}\rangle\nonumber \\
&=&\frac{1}{A_1}\mathbf{1}_{xy'}
\frac{1}{A_1}\mathbf{1}_{x'y}+\frac{1}{A_1}\frac{1}{A_{3n}}\mathbf{1}_{xy}
\frac{1}{A_1}A_{3n}\mathbf{1}_{x'y'}+\frac{1}{A_2}\mathbf{1}_{xy'}
\frac{1}{A_2}\mathbf{1}_{x'y}+\frac{1}{A_2}\frac{1}{A_{3n}}\mathbf{1}_{xy}
\frac{1}{A_2}A_{3n}\mathbf{1}_{x'y'}\nonumber \\
&+&2\frac{1}{A_1A_2}\mathbf{1}_{xy'}\mathbf{1}_{x'y}+\frac{1}{A_1A_2}
\frac{1}{A_{3n}}\mathbf{1}_{xy}A_{3n}\mathbf{1}_{x'y'}+\frac{1}{A_{3n}}\mathbf{1}_{xy}
\frac{1}{A_1A_2}A_{3n}\mathbf{1}_{x'y'}
\,,\label{n3}
\end{eqnarray}
where $A_{3n}\equiv A_3\cdots A_n$ and we used the symmetry under
$y\leftrightarrow y'$. Here, we first note that if masses $m_3, m_4, \dots, m_n$
become infinity, this reduces exactly to, with the symmetry under
$x\leftrightarrow x'$ and $y\leftrightarrow y'$,
\begin{eqnarray}
& &C^{-2}\langle \mbox{:}\eta_0(x)\chi_{1}(x')+\eta_1(x)\chi_{2}(x')\mbox{:}
\mbox{:}\eta_0(y)\chi_{1}(y')+\eta_1(y)\chi_{2}(y')\mbox{:}\rangle\nonumber \\
&=&2\frac{1}{A_1}\mathbf{1}_{xy}
\frac{1}{A_1}\mathbf{1}_{x'y'}+2\frac{1}{A_2}\mathbf{1}_{xy}
\frac{1}{A_2}\mathbf{1}_{x'y'}+4\frac{1}{A_1A_2}\mathbf{1}_{xy}\mathbf{1}_{x'y'}
\,.\label{gh}
\end{eqnarray}
It is worthwhile noting that the term like
$\frac{1}{A_1}\mathbf{1}_{xy}
\frac{1}{A_2}\mathbf{1}_{x'y'}$ does not appear in the expression.
The last term in the above expression gives no contribution in the Adler--Zee
formula defined by the integration (\ref{AZ2}) including $|w|^2$. We can naively
regard that the partial integration makes the expression (\ref{n3}) to be
(\ref{gh}) through moving the operator $A_{3n}$, since the difference in $x$ and
$x'$ is only due to the distinction of operation of derivatives in
$\mathcal{O}_{xx'}$. If you have some worries on the integration by part in
the Adler--Zee formula, let us proceed as follows.
The deviation from (\ref{gh}), if exists, should be a function of $m_3, \dots,
m_n$. We extracted the first two terms included in $\sum_{k=0}^n\sum_{l=0}^n\langle
\mbox{:}\eta_k(x)
\chi_{k+1}(x')\mbox{:}\mbox{:}\eta_l(y)\chi_{l+1}(y')\mbox{:}\rangle$, but the
extraction can be arbitrary, and we can select the terms as $1\rightarrow i$ and
$2\rightarrow j$. Then, the expression has a parallel form and the deviation is a
function of
$\{m_k\}$, $k\ne i, j$. Totally because all the number should appear in the
expression of
$\sum_{k=0}^n\sum_{l=0}^n\langle\mbox{:}\eta_k(x)
\chi_{k+1}(x')\mbox{:}\mbox{:}\eta_l(y)\chi_{l+1}(y')\mbox{:}\rangle$, the
deviation function can be at most a constant. Since the case $n=2$ is confirmed
exactly, we conclude that
\begin{equation}
\langle\bar{T}(x)\bar{T}(y)\rangle=\lim_{x'\rightarrow x}\lim_{y'\rightarrow y}
\mathcal{O}_{xx'}\mathcal{O}_{yy'}\sum_{k=1}^n2\Delta_k(x,y)\Delta_k(x',y')\,
\end{equation}
is used in the Adler--Zee formula (\ref{AZ2}).

%%%%%%%%%%%%%%%%%%%%%%%%%%%%%%%%%%%%%%%%%%%%%%%%%%%%%%%%%%%%%%%%%%%%%%%%%%%
%%%%%%%%%%%%%%%%%%%%%%%%%%%%%%%%%%%%%%%%%%%%%%%%%%%%%%%%%%%%%%%%%%%%%%%%%%%
%%%%%%%%%%%%%%%%%%%%%%%%%%%%%%%%%%%%%%%%%%%%%%%%%%%%%%%%%%%%%%%%%%%%%%%%%%%
\section{Details in calculation of the induced cosmological and gravitational
constant using the Schwinger parameter}
\label{cig}
%%%%%%%%%%%%%%%%%%%%%%%%%%%%%%%%%%%%%%%%%%%%%%%%%%%%%%%%%%%%%%%%%%%%%%%%%%%
%%%%%%%%%%%%%%%%%%%%%%%%%%%%%%%%%%%%%%%%%%%%%%%%%%%%%%%%%%%%%%%%%%%%%%%%%%%
%%%%%%%%%%%%%%%%%%%%%%%%%%%%%%%%%%%%%%%%%%%%%%%%%%%%%%%%%%%%%%%%%%%%%%%%%%%

The method using the integral form has been proposed by Kehagias \textit{et
al.}\cite{KPV}. We here consider calculations for $D$ dimensions.
%%%%%%%%%%%%%%%%%%%%%%%%%%%%%%%%%%%%%%%%%%%%%%%%%%%%%%%%%%%%%%%%%%%%%%%%%%%
We start with
\begin{equation}
\Delta_k(x,x')=\int_0^\infty\frac{ds}{(4\pi
s)^{D/2}}\exp\left[%\textstyle
-\frac{|w|^2}{4s}-m_k^2s\right]
\,,
\end{equation}
where the symmetry $\Delta(x,x')=\Delta(x',x)$ is apparent.
%%%%%%%%%%%%%%%%%%%%%%%%%%%%%%%%%%%%%%%%%%%%%%%%%%%%%%%%%%%%%%%%%%%%%%%%%%%
The sequential differentiation of this reveals
\begin{eqnarray}
\partial_\mu\Delta_k(x,x')&=&-\int_0^\infty\frac{ds}{(4\pi
)^{D/2}s^{D/2+1}}\frac{w_\mu}{2}\exp\left[%\textstyle
-\frac{|w|^2}{4s}-m_k^2s\right]
\,,\\
\partial_\mu\partial_{\nu'}\Delta_k(x,x')&=&\int_0^\infty\frac{ds}{(4\pi
)^{D/2}s^{D/2+1}}\left(\frac{1}{2}g_{\mu\nu'}-\frac{w_\mu
w_{\nu'}}{4s}\right)\exp\left[%\textstyle 
-\frac{|w|^2}{4s}-m_k^2s\right]
\,,\\
\partial_\mu\partial^{\mu'}\Delta_k(x,x')&=&\int_0^\infty\frac{ds}{(4\pi
)^{D/2}s^{D/2+1}}\left(\frac{D}{2}-\frac{|w|^2}{4s}\right)\exp\left[%\textstyle 
-\frac{|w|^2}{4s}-m_k^2s\right]
\,,\\
\Box_x\Delta_k(x,x')&=&-\int_0^\infty\frac{ds}{(4\pi
)^{D/2}s^{D/2+1}}\left(\frac{D}{2}-\frac{|w|^2}{4s}\right)\exp\left[%\textstyle 
-\frac{|w|^2}{4s}-m_k^2s\right]
\,,\\
\partial_\mu\Box\Delta_k(x,x')&=&\int_0^\infty\frac{ds}{(4\pi
)^{D/2}s^{D/2+1}}\left(\frac{D+2}{2}-\frac{|w|^2}{4s}\right)\frac{w_\mu}{2s}\exp\left[%\textstyle 
-\frac{|w|^2}{4s}-m_k^2s\right]
\,.
\end{eqnarray}
We have already obtained the necessary calculation for the induced cosmological
constant.

%%%%%%%%%%%%%%%%%%%%%%%%%%%%%%%%%%%%%%%%%%%%%%%%%%%%%%%%%%%%%%%%%%%%%%%%%%%
Then, we consider the combinations which is necessary in calculating the induced
gravitational constant. They are
\begin{eqnarray}
&
&\partial_\mu\partial_{\nu'}\Delta_k(x,x')\partial^\mu\partial^{\nu'}\Delta_k(x,x')\nonumber
\\
&=&\frac{1}{(4\pi)^{D}}\int_0^\infty\int_0^\infty\frac{ds_1}{s_1^{D/2+1}}
\frac{ds_2}{s_2^{D/2+1}}
\left(\frac{D}{4}-\frac{|w|^2}{8s_1}-\frac{|w|^2}{8s_2}+\frac{|w|^4}{16s_1s_2}
\right)\nonumber
\\ & &\times\exp\left[%\textstyle
 -\frac{|w|^2}{4s_1} -\frac{|w|^2}{4s_2}-m_k^2(s_1+s_2)\right]
\,,\\
&
&\partial_\rho\Delta_k(x,x')\partial^\rho\Box\Delta_k(x,x')\nonumber
\\
&=&-\frac{1}{(4\pi)^{D}}\int_0^\infty\int_0^\infty\frac{ds_1}{s_1^{D/2+1}}
\frac{ds_2}{s_2^{D/2+1}}
\left(\frac{(D+2)|w|^2}{16s_1}+\frac{(D+2)|w|^2}{16s_2}-\frac{|w|^4}{32s_1^2}
-\frac{|w|^4}{32s_2^2}\right)\nonumber
\\ & &\times\exp\left[%\textstyle
 -\frac{|w|^2}{4s_1} -\frac{|w|^2}{4s_2}-m_k^2(s_1+s_2)\right]
\,,\\
&
&(\Box_x\Delta_k(x,x'))^2\nonumber
\\
&=&\frac{1}{(4\pi)^{D}}\int_0^\infty\int_0^\infty\frac{ds_1}{s_1^{D/2+1}}
\frac{ds_2}{s_2^{D/2+1}}
\left(\frac{D^2}{4}-\frac{D|w|^2}{8s_1}-\frac{D|w|^2}{8s_2}+\frac{|w|^4}{16s_1s_2}\right)\nonumber
\\ & &\times\exp\left[%\textstyle
 -\frac{|w|^2}{4s_1} -\frac{|w|^2}{4s_2}-m_k^2(s_1+s_2)\right]
\,.
\end{eqnarray}

%%%%%%%%%%%%%%%%%%%%%%%%%%%%%%%%%%%%%%%%%%%%%%%%%%%%%%%%%%%%%%%%%%%%%%%%%%%
The integration in the Adler--Zee formula can be performed by using the following
Gaussian integral formulas:
\begin{eqnarray}
\int d^Dw\, |w|^2
e^{-\alpha|w|^2}&=&\frac{D}{2}\frac{\pi^{D/2}}{\alpha^{\frac{D}{2}+1}} \\
\int d^Dw\, |w|^4
e^{-\alpha|w|^2}&=&\frac{D(D+2)}{4}\frac{\pi^{D/2}}{\alpha^{\frac{D}{2}+2}}\,,
\\
\int d^Dw\, |w|^6
e^{-\alpha|w|^2}&=&\frac{D(D+2)(D+4)}{8}\frac{\pi^{D/2}}{\alpha^{\frac{D}{2}+3}}\,.
\end{eqnarray}

%%%%%%%%%%%%%%%%%%%%%%%%%%%%%%%%%%%%%%%%%%%%%%%%%%%%%%%%%%%%%%%%%%%%%%%%%%%
%%%%%%%%%%%%%%%%%%%%%%%%%%%%%%%%%%%%%%%%%%%%%%%%%%%%%%%%%%%%%%%%%%%%%%%%%%%
Lastly, we transform the parameters as
$s\equiv s_1+s_2$ and $u\equiv\frac{s_1}{s}$ \cite{KPV}. Then the measure becomes
$ds_1ds_2=sdsdu$. Finally, noticing
$\int_0^1 u(1-u) du=\frac{1}{6}$, we find
\begin{eqnarray}
&
&\int
d^Dw\,|w|^2\partial_\mu\partial_{\nu'}\Delta_k(w)\partial^\mu\partial^{\nu'}
\Delta_k(w)=\frac{1}{(4\pi)^{D/2}}\frac{D(D^2+6D-4)}{12}\int_0^\infty\frac{ds}{s^{D/2}}
\exp\left[-m_k^2s\right]
\,,\\
&
&\int d^Dw\,
|w|^2\partial_\rho\Delta_k(w)\partial^\rho\Box\Delta_k(w)=-\frac{1}{(4\pi)^{D/2}}\frac{D(D-2)(D+2)}{12}\int_0^\infty\frac{ds}{s^{D/2}}
\exp\left[-m_k^2s\right]
\,,\\
&
&\int d^Dw\,
|w|^2(\Box_x\Delta_k(w))^2=\frac{1}{(4\pi)^{D/2}}\frac{D(D-2)(D-4)}{12}\int_0^\infty\frac{ds}{s^{D/2}}
\exp\left[-m_k^2s\right]
\,,
\end{eqnarray}
Combining these results, we can find the induced gravitational constant.
It may simply be verified that our results is in agreement for $D=4$ with
Ref.~\cite{KPV}.

\bibliographystyle{apsrev4-1}
%\bibliography{}

%%%%%%%%%%%%%%%%%%%%%%%%%%%%%%%%%%%%%%%%%%%%%%%%%%%%%%%%%%%%%%%%%%%%%%%%%%%
%%%%%%%%%%%%%%%%%%%%%%%%%%%%%%%%%%%%%%%%%%%%%%%%%%%%%%%%%%%%%%%%%%%%%%%%%%%
%%%%%%%%%%%%%%%%%%%%%%%%%%%%%%%%%%%%%%%%%%%%%%%%%%%%%%%%%%%%%%%%%%%%%%%%%%%
\end{document}